\pdfoutput=1

\documentclass[aip,reprint,floatfix,final,nofootinbib]{revtex4-1} 

\usepackage{xcolor,graphicx,grffile,soul}
\usepackage[caption=false]{subfig} 
\usepackage{amssymb,amsfonts,amsmath} 
\usepackage[fixamsmath]{mathtools}
\usepackage[retainorgcmds]{IEEEtrantools} 
\usepackage{comment}
\usepackage[normalem]{ulem} 

\usepackage{ifdraft}  
\ifoptionfinal
  {\usepackage[textwidth=4cm,textsize=small,disable]{todonotes}}
  {\usepackage[letterpaper,hmargin={2cm,5cm},vmargin={2cm,2cm}]{geometry}
\usepackage[textwidth=4cm,textsize=small,disable]{todonotes}}

\usepackage[color]{showkeys}
\definecolor{refkey}{rgb}{0.5,0.,0.03}
\definecolor{labelkey}{rgb}{0.1,0.5,0.03} 

\usepackage[bookmarks,citecolor=blue,colorlinks,
  hyperfigures,linkcolor=blue,pdftex,pdfstartview=FitH,
  urlcolor=blue,pdfauthor={Elgar Pichler, Avi M.~Shapiro},
  pdftitle={Public Goods Games on Adaptive Coevolutionary Networks}]{hyperref}
\usepackage[framemethod=TikZ,roundcorner=5pt]{mdframed}

\graphicspath{{./figures/}}

\newcommand{\abs}[1]{\lvert#1\rvert}

\newcommand{\xtodo}[2][]{\todo[size=\small,#1]{#2}}

\newcommand{\AStodo}[2][]{\xtodo[color=green!40,#1]{#2}}
\newcommand{\EPtodo}[2][]{\xtodo[color=cyan!40,#1]{#2}}

\begin{document}

\title{Public Goods Games on Adaptive Coevolutionary Networks}

\author{Elgar Pichler}
\email[Electronic mail: ]{elgar.pichler@gmail.com}
\affiliation{Department of Chemistry and Chemical Biology, Northeastern University, Boston, MA 02115}

\author{Avi M.~Shapiro}
\email[Author to whom correspondence should be addressed. Electronic mail: ]{avi.m.shapiro@gmail.com}
\affiliation{Paulson School of Engineering and Applied Sciences, Harvard University, Cambridge, MA 02138}



\AStodo[inline,caption={new jargon}]{%
To describe network components, we currently use ``node'' and ``edge'' as in NetworkX. This mixes two conventions. Graph theory uses vertex/edge and CS/networks use node/link. I propose we use node/link since ``link'' is more descriptive of the interaction between players. This change would require remaking some plots in Section~\ref{sec:results}.
}

\begin{abstract}
Productive societies feature high levels of cooperation and strong connections between individuals.
Public Goods Games (PGGs) are frequently used to study the development of social connections and cooperative behavior in model societies. 
In such games, contributions to the public good are made only by cooperators, while all players, including defectors, can reap public goods benefits.
Classic results of game theory show that mutual defection, as opposed to 
cooperation, is the Nash Equilibrium of PGGs in well-mixed populations, where each player interacts with all others.
In this paper, we explore the coevolutionary dynamics of a low information public goods game on a network without spatial constraints in which players adapt to their environment in order to increase individual payoffs.
Players adapt by changing their strategies, either to \emph{cooperate} or to 
\emph{defect}, and by altering their social connections. 
We find that even if players do not know other players' strategies and connectivity, cooperation can arise and persist despite large short-term fluctuations.
\end{abstract}


\maketitle

\begin{quotation}
Evolutionary models of cooperation have been actively studied by biologists, physicists, economists and mathematicians. 
Early game theoretic models involving two players, such as the Prisoner's Dilemma game, have been generalized to multiplayer games where players, or agents, interact with all others as cooperators or defectors.
Iterative games allow players to change strategies, often with knowledge of neighbor strategies and payoffs.
Subsequently, spatial structure has been incorporated into multiplayer games by arranging players on simple lattices defining local interaction partners.
Since social connections in applications are often not limited by space, complex networks have become the arena for numerous games investigating the emergence and stability of cooperation.
Games on networks with diverse topologies have illuminated the importance of network structure on dynamics, yet most models assume a static network topology. 
We study Public Goods Games on evolving complex networks to model the realistic setting where social structure varies along with individual strategy changes.
We find parameter regimes where cooperation is favored and players become highly connected, all without knowing the strategies of neighboring players.
\end{quotation}

\section{Introduction}
\label{sec:introduction}

The emergence of social cohesion, cooperation and connectivity between individuals, are general features essential for the overall functioning of a society.
Successful societies tend to have high levels of cooperation and social cohesion\cite{Nowak_2006, Wakano_2009} while the opposite is true of less functional societies.\cite{Nowak_2006}
It has been shown that the interaction pattern between individuals or the topological structure of social groups plays a significant role in the evolution of cooperative behavior, although conclusions vary.
Whether network heterogeneity favors cooperation is unclear. 
For prisoner dilemma games, regular graphs support cooperation the most  while scale-free graphs do not support cooperation at all,\cite{Konno_2011} however it is hypothesized that this may depend on the specific rules of games.
Indeed, scale-free graphs are shown to support cooperation in similar games.~\cite{Santos_2006b,Santos_2006,Santos_2008}
In most studies, including these, the network structure is static throughout the evolutionary game, unlike in real societies.

To study the emergence of cooperation in a more general and realistic setting, we consider a game theoretic model of a group of interacting individuals, a repeated public goods game (PGG) involving mobile players with two different behavioral strategies: cooperation and defection. 
A PGG generalizes the Prisoner's Dilemma to more than two players and allows for greater diversity in system evolution.\cite{Szabo_2007,GomezGardenes_2011,Santos_2008,Archetti_2012,Axelrod_1981}

In a PGG, cooperators contribute a certain amount to the public good and defectors do not, yet all players receive an equal share of the sum of contributions, multiplied by a synergy factor. 
If every player cooperates, then everyone benefits,  
and if every player defects, then no one benefits.  
Clearly, a single player is better off when many or all other game participants cooperate. 
However, an individual who defects still receives the group benefits without any cost. 
This so-called \emph{free-rider problem},\cite{Hardin_1968} 
leads to a situation in which no one cooperates. 
Still, cooperation persists in real societies.

In a well-mixed population, where every player interacts with every other player, the Nash equilibrium of PGGs is total defection, leading to the \emph{tragedy of the commons},\cite{Hardin_1968} in which public resources are exhausted.
Spatially restricted versions of PGGs have been studied, typically on simple lattices with nearest neighbor interactions,\cite{Roca_2011} in order to demonstrate how cooperation can flourish in more complicated yet realistic systems.
Motivated by applications to peer-to-peer, mobile, and vehicular networks, we generalize such studies by allowing varying connections between arbitrary players on arbitrary nodes in a non-planar network topology.
Allowing for the network topology to coevolve with player strategy has been studied using the Prisoner's Dilemma game\cite{Zimmermann_2004,Biely_2007,Ebel_2002a} and various other symmetric two-player games.\cite{Pacheco_2006,Pacheco_2006b}

Models of strategy and network evolution often follow a random \emph{satisfying} dynamic\cite{Ebel_2002a,Simon_1955} 
reminiscent of biological (genetic) evolution, or an \emph{imitation} dynamic\cite{Kuperman_2001,Zimmermann_2004} reminiscent of social evolution. 
In both cases, a player's decisions to change strategies or connections is dependent on knowledge of neighbors' strategies and payoffs, a high information requirement which often does lead to high levels of cooperation. 
However, even among knowledgeable humans, experiments have shown individuals often do not always imitate the best neighbor strategy.\cite{Traulsen_2010}
In order to create a minimal model, we consider a low information setting in which player decisions depend only on individual payoffs. 
Despite lacking mechanisms such as punishment\cite{Brandt_2003,Helbing_2010} or reputation effects,\cite{Brandt_2003} which are known to promote cooperation, we observe 
both high levels of cooperation and connectivity for a range of model parameters while the adaptive network develops a highly peaked degree distribution.

\section{Public Goods Game}
\label{sec:pgg}

\subsection{PGG in well-mixed populations}\label{sec:mixed_pgg}
In a PGG, cooperators contribute a certain amount to the public good and defectors do not. 
All players then receive an equal share of the sum of all contributions multiplied by a \emph{synergy} factor, $r\ge1$, accounting for non-zero-sum benefits of cooperation.
Given $n$ mutually interacting players including $n_{\text{\tiny C}}$ cooperators, we can write the payoffs for cooperators and defectors,%
\begin{equation} \label{eq:payoffs}
	\pi_{\text{\tiny C}} = r c\frac{n_{\text{\tiny C}}}{n} - c, \quad 
	\pi_{\text{\tiny D}} = r c\frac{n_{\text{\tiny C}}}{n},
\end{equation}%
\noindent where $c$ is the fixed \emph{cost} of cooperation per game.  
In common formulations, $b=rc$ is the \emph{benefit} of cooperation so that synergy, $r=b/c$, is the benefit-cost ratio.\cite{Nowak_2006,Ohtsuki_2006,Ohtsuki_2006b,Konno_2011}
Without loss of generality, we take $c=1$. 
Other variations of PGGs have
a fixed cost per individual, with an equal fraction contributed to each game the individual plays.
This variation is only quantitatively distinct in the heterogeneous populations discussed below in Section~\ref{sec:network_pgg}, yet is often more supportive of cooperation than the one considered here.\cite{Santos_2008,Szabo_2007}  
Our modeling decisions have been guided by the avoidance of common mechanisms to induce widespread cooperation such as those mentioned in Section~\ref{sec:introduction}.

Cooperators earn a lower payoff than defectors since Equations~\eqref{eq:payoffs} clearly imply $\pi_{\text{\tiny C}} < \pi_{\text{\tiny D}}$. 
If all $n$ players are defectors, their payoffs are zero. 
If one player is instead a cooperator, its payoff decreases and becomes negative if $r<n$. 
Hence, an individual defector has no incentive to unilaterally switch to become a cooperator. 
The Nash Equilibrium in a PGG is thus total defection.
In repeated games, as we consider here, players may change strategies after each round based on their payoffs. 
In a well-mixed population with homogeneous social structure, when players  seek to maximize their payoffs, the dynamics lead to total defection, as in the single round game.

To investigate a mechanism which promotes cooperation, one may relax the well-mixed assumption and include a spatial interaction structure so that individuals only play with a subset of the population.
Spatial games often take place on simple lattices.\cite{Nowak_1994} 
To generalize such local interactions, this paper focuses on a PGG on a dynamic, complex network described in Section~\ref{sec:network_pgg}.

\subsection{PGG on a complex network}\label{sec:network_pgg}
\begin{figure}[!hbt]
  \includegraphics[width=.7\linewidth]{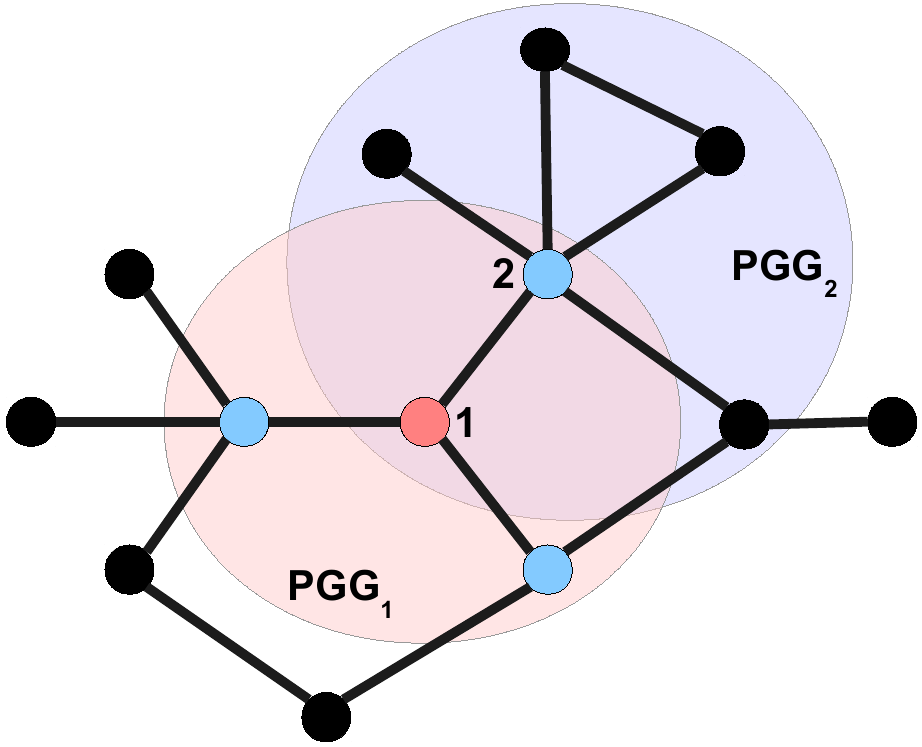}
  \caption{Example network of public goods games\label{fig:pgg}}
\end{figure}

On a network or graph consisting of nodes and edges, a separate game is played in a neighborhood centered on each player.
A neighborhood is composed of a central player and all neighbors, the players connected to the central player by a single edge.
A player's total payoff is accumulated from each game played. 
So a highly connected player can earn payoffs from many games.
For example, in Figure~\ref{fig:pgg},  a player at node 1 has three neighbors and thus collects a total payoff from four separate games (only two are shown).
\AStodo[noinline]{Add 2 more circles for other games including node 1. See \cite{Santos_2008} for example.} 
A player without any connected neighbors does not play any games.

In this network setting, each player can be connected to any other, so the number of players in a single game is only bounded by the total network size.
In contrast, on a 2D regular square lattice, players are only connected to nearest neighbors (either at most 4 for a von Neumann or 8 for a Moore neighborhood\cite{Roca_2011}). 
Similarly, location (and thus, neighbor) changes for a player are usually limited to spaces within a certain distance in such cellular automata.
With such restrictions, an unsatisfied player may not be able to change its location or connections to other players sufficiently to reflect its level of dissatisfaction.

Players make separate decisions about strategy and connectivity and each decision creates its own dilemma.
Profit-maximizing players should all choose the Nash equilibrium strategy of defection (if $r<n$), which leads to zero payoff for all players, clearly the worst possible state in terms of total wealth. 
This is the classic social dilemma which also exists in the well-mixed case.
On a network, it is also advantageous to participate in a large number of games with many cooperators.
However, this runs the risk of being exploited by many defectors. 
This is the secondary dilemma due to the irregular distribution of edges and node strategies throughout the dynamic network.
In Section~\ref{sec:model}, we introduce repeated PGG update and decision rules which overcome these dilemmas.

\section{Coevolutionary Dynamics}
\label{sec:model}
The following dynamical rules closely follow Roca et al.\cite{Roca_2011} 
The behavior of the players is driven by the individual player's satisfaction. 
Based on the level of satisfaction, a player may change strategy and/or connectivity to other players as detailed in the following sections.

\subsection{State variables}
At time step $t$, player $i$ may change strategy and/or neighbors (network plasticity) with independent probabilities based on its satisfaction,
\begin{equation}\label{eq:satis}
s_{i}(t) = \pi_{i}(t)-a_{i}(t)+\eta_{i}(t),
\end{equation}
where $\pi_{i}$ is the net payoff from all games played by player $i$, $\eta_{i}$ is a Gaussian noise, and $a_{i}$ is \emph{aspiration}, defined 
\begin{align}
a_{i}(t) 
  &= \alpha\, \pi_{i,\text{max}}(t) + \left( 1 - \alpha \right) \pi_{i,\text{min}}(t) \nonumber\\
  &= \pi_{i,\text{min}}(t) + \alpha \bigl( \pi_{i,\text{max}}(t) -\pi_{i,\text{min}}(t) \bigr), \label{eq:aspir}
\end{align}
which depends on \emph{greediness} $0\le\alpha\le1$, another important control parameter. 
Aspiration is a convex combination of functions $\pi_{i,\text{max/min}}$ which reflect (finite) memory of past max/min payoffs.
A habituation effect is modeled by defining the update equations
%
\begin{align}
  \phantom{\pi_{i,\text{max}}(t+1) =}
  &\begin{aligned}\label{eq:max_update}
    \mathllap{\pi_{i,\text{max}}(t+1) =}\; \pi_{i}(t) + (1-\mu) 
      &\left( \pi_{i,\text{max}}(t)-\pi_{i}(t) \right) \\
      &H\bigl( \pi_{i,\text{max}}(t) -\pi_{i}(t)\bigr),
  \end{aligned}\\
  &\begin{aligned}\label{eq:min_update}
   \mathllap{\pi_{i,\text{min}}(t+1) =}\; \pi_{i}(t) - (1-\mu)
     &\left( \pi_{i}(t) - \pi_{i,\text{min}}(t) \right) \\
     &H\bigl( \pi_{i}(t)-\pi_{i,\text{min}}(t) \bigr),
 \end{aligned}
\end{align}
where $H$ is the Heaviside step function.
The quantity $1-\mu$ describes a fractional memory length. Our choice, $\mu=0.01$, implies a long but finite memory. The noise term $\eta_{i}(t)$ is drawn from a Gaussian distribution with mean zero and standard deviation $0.1$.

\subsection{System Evolution}
At each time step, an unsatisfied player ($s_{i}<0$) is allowed to make any or all of three updates: change strategy, add an edge to a random player (excluding self-edges and multiple edges), or remove an edge to a random neighbor. The outcomes of each possible update are based on three independent trials with probability
\begin{equation}\label{eq:prob}
	p_{i}(t) = \tanh (\abs{s_{i}}/k) H(-s_{i}).
\end{equation}
The scaling factor, $k=8$, as in Ref.~\onlinecite{Roca_2011}, and ensures a range of probabilities so that very unsatisfied players are more likely to change their state than slightly unsatisfied players. 
The Heaviside function ensures that players with non-negative satisfaction do not change their strategies or neighbors.
While this probability function is continuous in satisfaction, it changes quickly as satisfaction crosses zero and is thus very sensitive.
A player's satisfaction and change probability depend strongly on the current network topology. 
To allow each player to change neighbors and/or strategy based on the same network state, updating is performed synchronously.

\section{Results}
\label{sec:results}
\subsection{Initial Conditions}
To test whether cooperation can develop in a non-cooperative environment, we assume all players are initially defectors and allow cooperation to emerge stochastically due to the noise in Equation~\eqref{eq:satis}.
Our main results are simulated using an initial network with no edges. 
Remarkably, a variety of other initial networks, including small world and scale-free networks, result in very similar long time behavior.

While we have conducted simulations using various network sizes and parameters, we present results on PGGs consisting of $1000$ players and focus on the effects of the two parameters \emph{synergy}, $r$, and \emph{greediness}, $\alpha$ on cooperation and network structure. 

\subsection{PGG Dynamics/Evolution}
\subsubsection{Single simulation}
\begin{figure}[!htb] 
  \includegraphics[scale=1]{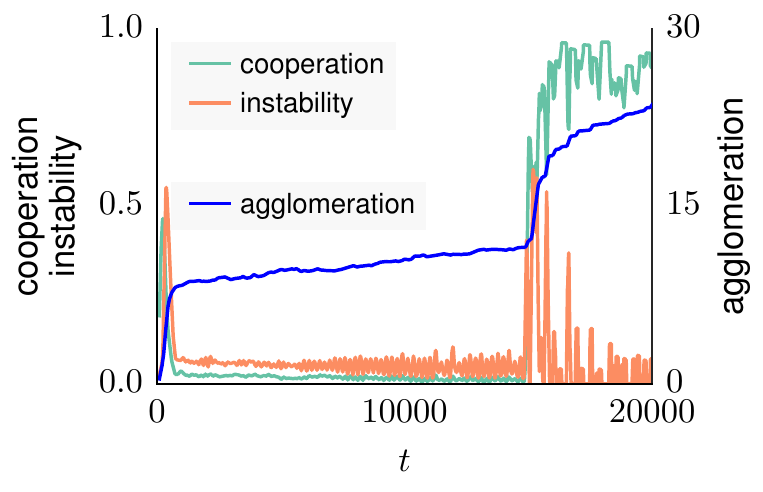} 
  \caption{Typical simulation run that favors cooperation. The network has 1000 players, initially all defectors with zero edges. Game parameters are synergy, $r=4$, and greediness, $\alpha=0.65$. Time series are averaged over a moving window of $100$ time steps.}
  \label{fig:3fields}
\end{figure}

The model evolution is captured by the variables \emph{cooperation}, the cooperator fraction%
, \emph{instability}, the fraction of players changing strategies or neighbors, and \emph{agglomeration}%
, the average node degree.%
\footnote{While the clustering coefficient might be a more appropriate measure on a network, results are qualitatively similar to those of the simpler measure average degree.}
Time series for these variables are shown in Figure~\ref{fig:3fields}, where $r=4$ and $\alpha=0.65$.\AStodo[noinline]{Check if B/W friendly. IT'S NOT. Default prettyplotlib colors used are terrible in grayscale.}{}
To observe overall trends through the stochasticity, time series show data averaged over a moving window of 100 time steps.

We observe several characteristic periods of behavior. 
Initially, many edges are added and sustained, increasing average node degree rapidly from zero while there is a short lived spike in cooperation. 
Second, agglomeration increases slowly while cooperation remains low but with growing fluctuation. 
Third, cooperation jumps up and agglomeration follows. 
Lastly, cooperation remains high while agglomeration grows slowly.

\begin{figure}[!htb] 
  \includegraphics[scale=1]{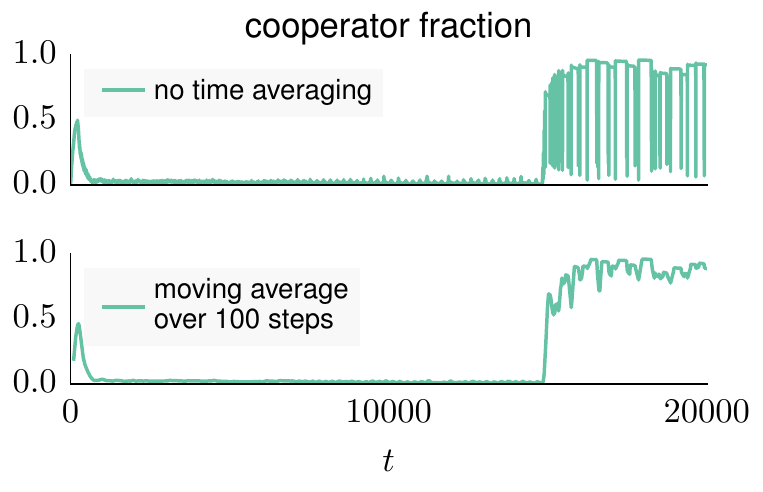} 
  \caption{Large amplitude fluctations in cooperation time series from simulation shown in Figure~\ref{fig:3fields}. (upper) Short but large variations in cooperator fraction are present in the raw data. (lower) Averaging over a moving window of 100 time steps reveals overall trend while smoothing high frequency oscillations.}
  \label{fig:stoch_coop}
\end{figure}

\begin{figure}[!htb] 
  \includegraphics[scale=1]{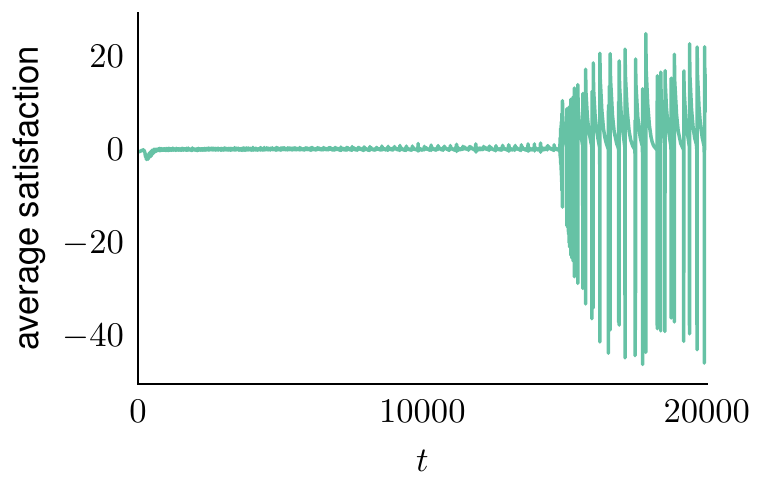} 
  \caption{Network averaged satisfaction time series. The positive spikes and subsequent exponential decay cause the oscillations in cooperation in Figure~\ref{fig:stoch_coop}.
  Time series is from the same simulation as used in Figures~\ref{fig:3fields} and~\ref{fig:stoch_coop} and is shown without moving window averaging.}
  \label{fig:satis}
\end{figure}

Stochasticity in the system arises from the aspiration noise in Equation~\eqref{eq:satis} and causes large amplitude cooperation fluctuations shown in Figure~\ref{fig:stoch_coop} and Appendix~\ref{app:oscillations}.
The upper plot in Figure~\ref{fig:stoch_coop} shows the original highly oscillatory cooperator time series while the lower is averaged over a sliding window of 100 time steps, as in Figure~\ref{fig:3fields}. 
The amplitude of oscillations in the raw data reveal a high level of strategy synchronization between players even though players make independent decisions.
These fluctuations and resulting synchronization are due to the habituation effect in satisfaction modeled in equations~\eqref{eq:satis}, \eqref{eq:aspir}, \eqref{eq:max_update}, and~\eqref{eq:min_update}. 
The evolution of satisfaction is shown in Figure~\ref{fig:satis} without window averaging. 
Due to Equations~\eqref{eq:aspir}, \eqref{eq:max_update}, and~\eqref{eq:min_update}, unless recent payoffs are higher than $\pi_{i,\text{max}}$, satisfaction will decrease toward zero.
At late times, after the onset of high cooperation, we do observe positive average satisfaction decaying to zero, then a large number of strategy changes toward defection dropping satisfaction far below zero.
This induces strategies to revert back toward cooperation and satisfaction resets at a positive value.

PGG evolution only occurs when satisfaction decreases below zero, i.e.~when aspiration nears payoff.
This is an imposed feature of the model, described by the equations in Section~\ref{sec:model}, yet, as we have seen, has interesting consequences. 
For long times, when high cooperation and agglomeration have been reached, the network is still subject to intermittent cooperation crashes which occur with some regularity.  
The time scale of such crashes should depend on $\mu$ and $\alpha$. 
Here we make that dependency explicit under some mild assumptions and show that higher $\mu$ and $\alpha$ increase the frequency of cooperation collapses.
 
Consider a satisfied player $i$ whose neighbors and their strategies are not changing at a given time step $t_{0}$. 
Then, for some time later, the player's payoff is constant, $\pi_{i}(t)\approx\pi_{i}(t_{0})$ for $t_{0}\le t\le t_{1}$. 
Assume also the strict bounds $\pi_{i,\text{min}}(t) < \pi_{i}(t) < \pi_{i,\text{max}}(t)$. 
The relaxation of the maximum and minimum payoffs toward the current (constant) payoff occurs at the same rate as the relaxation of satisfaction to the size of the noise. 
Given the stated assumptions on payoff, it is shown in Appendix~\ref{app:habituation} that
\begin{equation}\label{eq:satisfaction_decay}
s_{i}(t) \approx K_{i}(\alpha) e^{-\mu(t-t_{0})} + \eta_{i}(t),\quad t_{0}\le t\le t_{1},
\end{equation}
where 
\begin{equation}
  \begin{aligned}
    K_{i}(\alpha)
      &=\left[\pi_{i}(t_{0})- \pi_{i,\text{min}}(t_{0}) \right] \\
      &\quad {}- \alpha \left[ \pi_{i,\text{max}}(t_{0}) - \pi_{i,\text{min}}(t_{0}) \right]
  \end{aligned}
\end{equation}
is a linearly decreasing function of $\alpha$.

The behavior of satisfaction shown in Equation~\ref{eq:satisfaction_decay} leads to several important conclusions regarding the effects of the memory and greediness parameters. 
Recall, the faster satisfaction decays to zero, the sooner a collapse in cooperation occurs.
For short memory (high $\mu$), satisfaction decreases toward zero faster, increasing the frequency of cooperation collapse.

If $\mu$ is fixed, then what controls the duration of a period of quasi-stability in cooperation, as seen in the upper plot in Figure~\ref{fig:stoch_coop}, is the initial amplitude of the satisfaction decay, $K_{i}(\alpha)$. 
A lower amplitude reduces the time required for a player to become dissatisfied and shortens the interval between huge collapses in cooperation, making such collapses more frequent.
Three factors contribute to a lower amplitude. 
Clearly, and most importantly, the expression for $K_{i}(\alpha)$ shows the amplitude decreases linearly with greediness.
Additionally, if the payoff is close to its minimum or if the difference between the maximum and minimum payoffs is large, the amplitude is also diminished.
These variations and the random noise are responsible for the irregular timing of cooperation collapses even for a fixed $\alpha$ and $\mu$.

Notice that synergy, $r$, does not appear explicitly in Equation~\eqref{eq:satisfaction_decay}, but actually scales all of the payoffs in $K_{i}(\alpha)$ due to Equation~\eqref{eq:payoffs}.  
In other words, higher synergy should increase satisfaction, as expected.

\subsubsection{Multiple simulations}
\begin{figure}[!htb] 
  \includegraphics[scale=1]{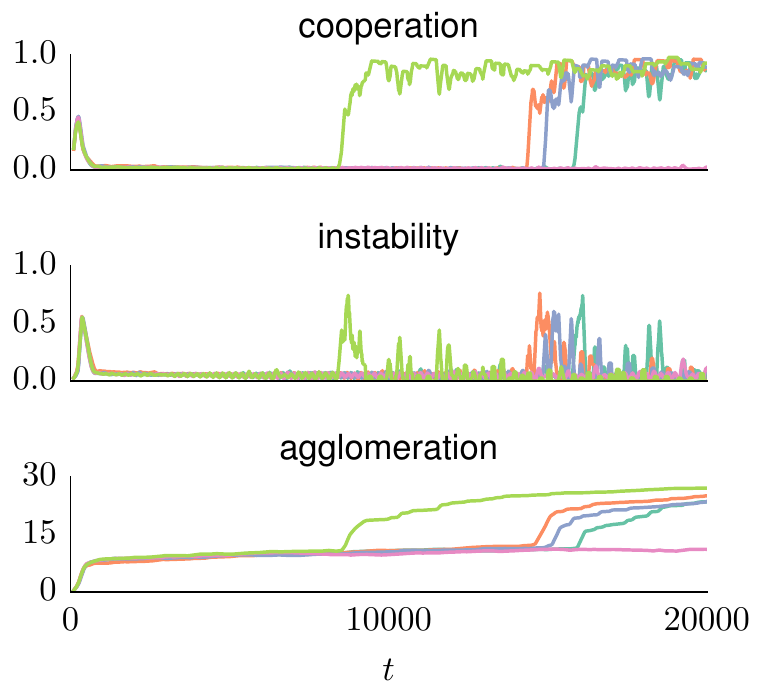} 
  \caption{Stochasticity in onset of cooperation. Multiple simulations for the parameters in Figure~\ref{fig:3fields} show similar behavior but shifted in time due to the stochasticity of the system.}
  \label{fig:stoch}
\end{figure}

Several simulations for each parameter pair reveal consistently similar behavior, although shifted in time; see Figure~\ref{fig:stoch}.
%
This is one example of the system's strong sensitivity to small perturbations due to random edge attachment and the sharp difference in the probability of state changes for satisfactions near zero.

\begin{figure*}[!htb]
  \includegraphics[scale=1]{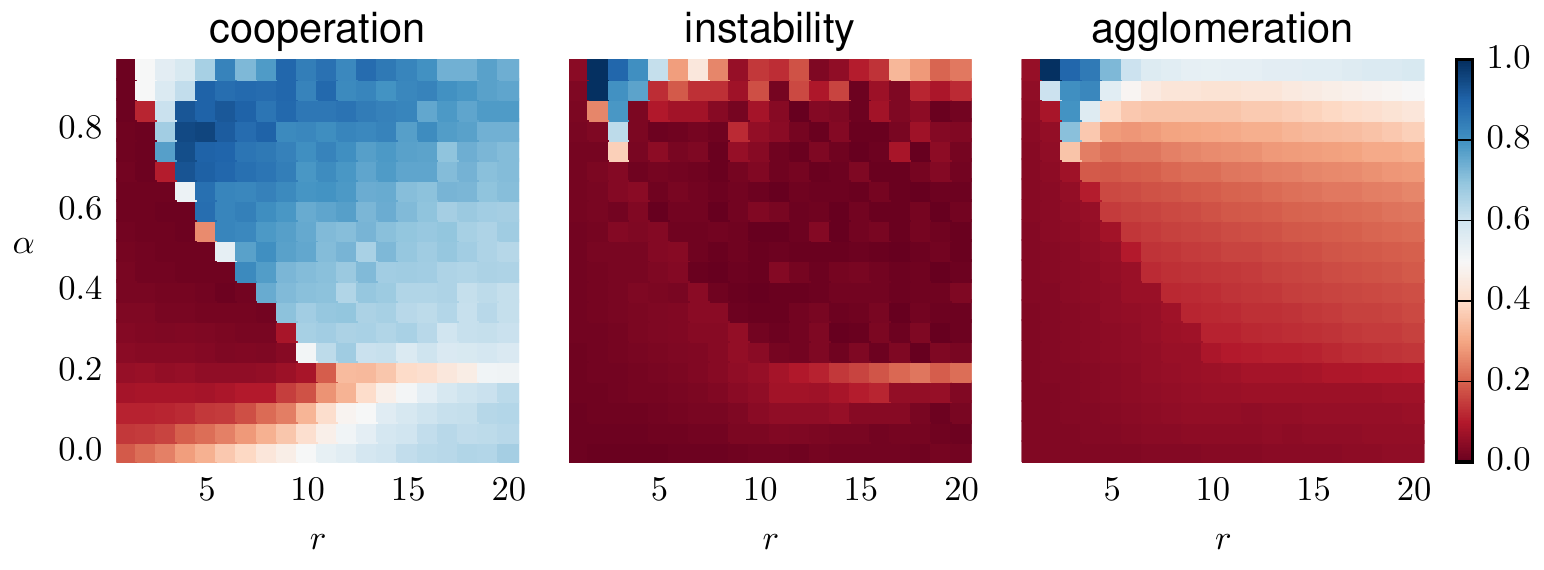}
  \caption{Phase diagrams depicting the fraction of cooperators, fraction of strategy and/or edge changers, and scaled average node degree (agglomeration) of PGGs as functions of synergy, $r$, and greediness, $\alpha$. Values are computed for PGGs with 1000 players averaged over the last 100 of 20,000 time steps. 
  \label{fig:PhaseDiagram}}
\end{figure*}

Despite strong time dependence of cooperation, long time averages show, similarly to the 2D lattice setting in Ref.~\onlinecite{Roca_2011}, the existence of parameter regions which result in both high and low cooperativity after sufficient rounds of PGGs.
However, Figure~\ref{fig:PhaseDiagram} shows that the transition between these regions is sharper in the network setting.

This phase diagram shows greediness encourages cooperation, especially for low synergy where the cost to cooperate is highest.
When synergy is at least 10, high cooperation results for all levels of greediness.
Remarkably, given Equations~\eqref{eq:payoffs}, the highest cooperation levels do not occur at highest synergy. 
Low instability throughout most of phase space shows the cooperation levels are nearly stationary at the end of simulations. 

\subsection{Network Topology}
In order to study the sharp divide between cooperator and defector dominated regions in phase space seen in Figure~\ref{fig:PhaseDiagram}, we investigate the long time network degree distribution and time dependent edge placement.

\subsubsection{Degree distribution}
\begin{figure}[!htb] 
  \includegraphics[scale=1,clip=true,trim=0 0 0 0]{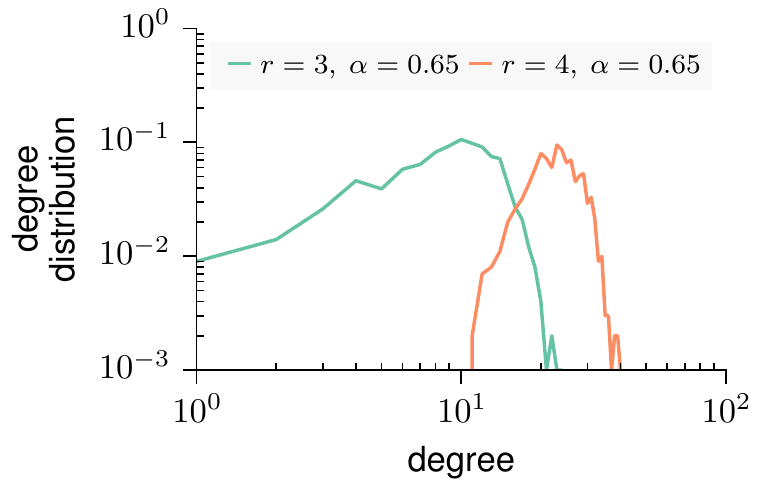} 
  \caption{Degree distribution for a $1000$ node network at $t=20000$. The $r=3$ ($r=4$) distribution corresponds to a low (high) cooperation long term state.}
  \label{fig:degree-dist}
\end{figure}

The network topology evolves during the simulation and a player can have between 0 and $n-1$ edges.
As players add connections to others, their node degree increases.
The long term degree distribution is surprisingly narrow and peaked for parameters which exhibit high cooperation, such as for $r=4$, $\alpha=0.65$ shown in Figure~\ref{fig:degree-dist}.
Note the absence of high degree nodes or hubs.
This distribution is suggestive of an Erd\H{o}s-R\'{e}nyi random graph.\cite{Erdos_1960} In fact, this PGG could be used as a network growth mechanism.

\subsubsection{Edge dynamics}
While full network topology as shown by degree distribution is clearly quite different for parameters exhibiting high and low cooperation, it is valuable to look closer at the connections of cooperators and defectors separately over time.
In particular, we would like to determine whether cooperators are located at nonrandom nodes in the graph and how their statistics differ from defectors. 

\begin{figure}[!htbp] 
  \includegraphics[scale=1]{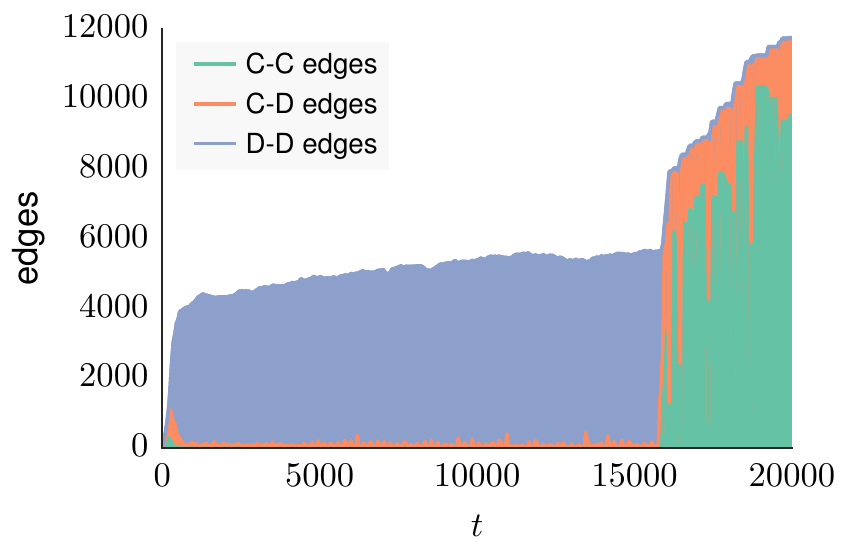} 
  \caption{Stacked bar plot for edges between two cooperators (C-C), a cooperator and a defector (C-D), and two defectors (D-D). Note the resemblance of the total edges curve to the agglomeration plot from the same simulation in Figure~\ref{fig:3fields}. $r=4$ and $\alpha=0.65$.}
  \label{fig:edgetypes}
\end{figure}

We begin by considering the dynamics of edges being added to each type of player for the case $r=4$ and $\alpha=0.65$, parameters near the sharp transition in Figure~\ref{fig:PhaseDiagram}.
The edge dynamics in Figure~\ref{fig:edgetypes} show growth in edges involving cooperators but not to the extent that cooperators are accumulating edges preferentially.
While the total number of edges increases, the distribution of edge types shows a strong and meaningful relationship to the number of cooperators, which is also changing.  
In fact, the number of C-C, C-D, and D-D edges varies with cooperation as would be expected if edges were placed randomly.

\begin{figure}[htb]  
  \subfloat[]{\includegraphics[]{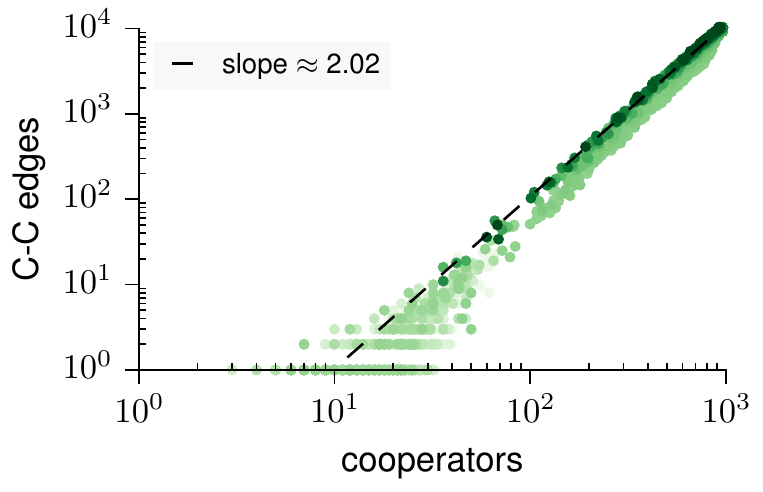}\qquad \label{fig:edgesCC}}\\
  \subfloat[]{\includegraphics[]{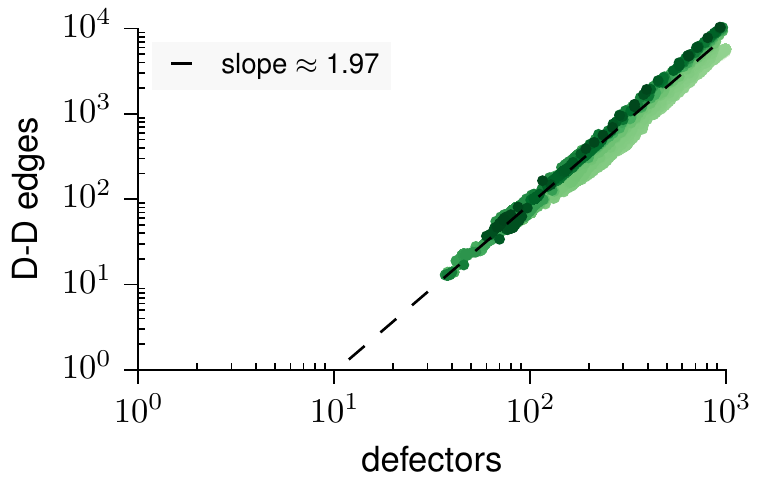}\qquad \label{fig:edgesDD}}  
  \caption{Edge counts between players of like strategy. The passage of time is indicated by the darkening of points with data taken from the simulation used in Figure~\ref{fig:3fields}. The data range plotted is centered at the onset of high cooperation, near $t=15000$. $r=4$ and $\alpha=0.65$.}
  \label{fig:intraedges}
\end{figure}

\begin{figure}[htb] 
  {\includegraphics[]{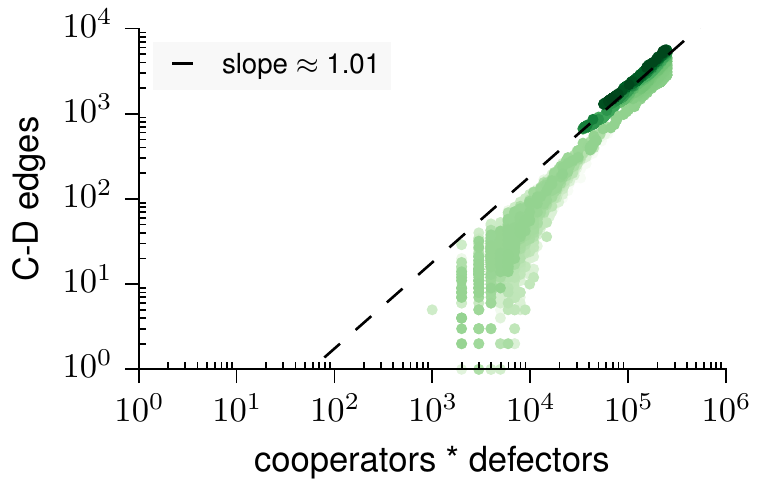}\qquad}
  \caption{Edge counts between cooperators and defectors. Parameters are as in Figure~\ref{fig:intraedges}.}
  \label{fig:edgesCD}
\end{figure}

First, consider a complete graph $K_{N}$ with $N=N_{\text{\tiny C}}+N_{\text{\tiny D}}$ nodes of cooperators and defectors. 
Then we can count the number of edges connecting any two nodes, $M$, two cooperators, $M_{\text{\tiny CC}}$, a cooperator and a defector, $M_{\text{\tiny CD}}$, and two defectors, $M_{\text{\tiny DD}}$, as follows
\begin{gather}\label{eq:scaling}
\begin{aligned}
  M &= \begin{pmatrix}N\\2\end{pmatrix}= \frac{1}{2}N(N-1),\\
  M_{\text{\tiny CC}} &= \begin{pmatrix}N_{\text{\tiny C}}\\2\end{pmatrix}= \frac{1}{2}N_{\text{\tiny C}}(N_{\text{\tiny C}}-1),\\
  M_{\text{\tiny CD}} &= N_{\text{\tiny C}}N_{\text{\tiny D}}= N_{\text{\tiny C}}(N-N_{\text{\tiny C}}),\\
  M_{\text{\tiny DD}} &= \begin{pmatrix}N_{\text{\tiny D}}\\2\end{pmatrix}= \frac{1}{2}N_{\text{\tiny D}}(N_{\text{\tiny D}}-1).
\end{aligned}
\end{gather}
In an Erd\H{o}s-R\'{e}nyi random graph with $N$ nodes and edge probability $p\in(0,1)$, the edge counts would all be multiplied by $p$ on average. 
In other words, the scaling of edges versus nodes would be the same as in Equations~\eqref{eq:scaling}. 
The log-log plots of PGG graph data in Figure~\ref{fig:edgesCC} show $M_{\text{\tiny CC}}$ and $M_{\text{\tiny DD}}$ are quadratic in $N_{\text{\tiny C}}$ and $N_{\text{\tiny D}}$, respectively. 
Likewise, Figure~\ref{fig:edgesCD} shows $M_{\text{\tiny CD}}$ is linear in the product $N_{\text{\tiny C}}N_{\text{\tiny D}}$.
In the plots, points darken as time advances and a restricted data range is chosen centered at the onset of high cooperation, near $t=15000$.
Indeed, we observe in the PGG, long time edge scaling consistent with an
Erd\H{o}s-R\'{e}nyi random graph.

\subsubsection{Cooperator clustering}
\begin{figure}[htb]
\flushleft
  \includegraphics[width=\linewidth]{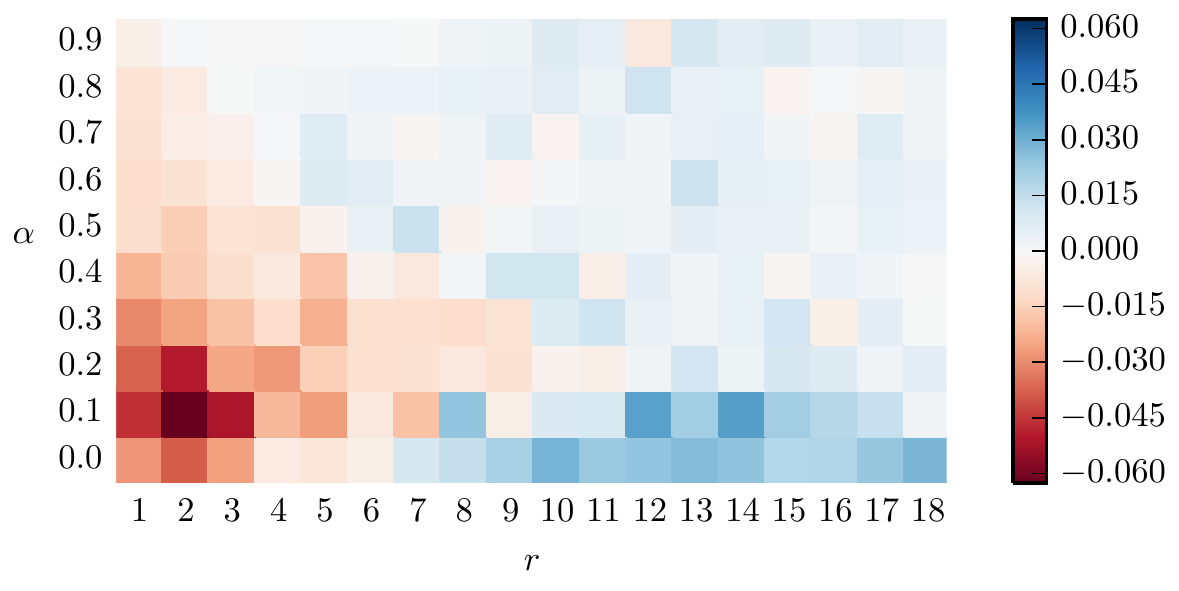}
  \caption{Deviation of neighborhood cooperator fraction from full network cooperator fraction, $\langle(n_{\text{\tiny C}}-1)/(n_{\text{\tiny C}}-1+n_{\text{\tiny D}})\rangle - (N_{\text{\tiny C}}-1)/(N_{\text{\tiny C}}-1+N_{\text{\tiny D}})$. Parameters are as in Figure~\ref{fig:intraedges}.}
  \label{fig:enrichment}
\end{figure}

With edges suggesting the network is evolving into a random graph, it is not clear whether cooperators and defectors occupy completely random positions within the network.
We would like to determine whether strategy statistics for a player's immediate neighbors differ significantly from statistics of the whole network.
Given a network, we can simply calculate if cooperators have a higher fraction of cooperating neighbors, on average, $\langle(n_{\text{\tiny C}}-1)/(n_{\text{\tiny C}}-1+n_{\text{\tiny D}})\rangle$, than the fraction of cooperators in the whole network, $(N_{\text{\tiny C}}-1)/(N_{\text{\tiny C}}-1+N_{\text{\tiny D}})$.
If this is the case, then cooperators are showing evidence of clustering. (Note that we subtract one in the above ratios to omit a central cooperator from the counts.)
The difference between these two ratios is shown for various greediness and synergy values in the heatmap in Figure~\ref{fig:enrichment}. 
One simulation for each parameter pair has been used rather than an ensemble average. 
Positive values indicate neighborhoods having more cooperators than would be expected if they were uniformly distributed over the network.

Figure~\ref{fig:enrichment} shows that there are parameter regions in which some local cooperator clustering is present. 
These regions resemble the regions of high cooperation in the leftmost plot in Figure~\ref{fig:PhaseDiagram}. 
In other words, when cooperation flourishes, cooperators are taking advantage of network reciprocity.\cite{Nowak_2006,Ohtsuki_2006,Konno_2011} 
However, the magnitude of this effect is very small, since the values in Figure~\ref{fig:enrichment} are generally very close to zero. 
To get a sense of scale, imagine that 1\% of the players were to become cooperators; then the values in the heatmap would decrease by approximately $0.010$, an amount greater than the magnitude of most of the current values.

\section{Summary}
\label{sec:summary}

The behavior of a spatially constrained, low information (no exchange of strategy, position, or connectivity) repeated PGG was previously analyzed by Roca et al.\cite{Roca_2011} with a cellular automaton model on a 2D square lattice. 
In such models, players can only interact with their nearest neighbors (8 in the case of a Moore neighborhood on a planar square lattice). 
The most important finding was that cooperativity arises and persists in such systems for certain parameter regions.

In this paper, we have investigated the behavior of the same low information PGG system with spatial restrictions removed using a network model. 
Here, players interact with others without regard to any distance metric. 
Single game sizes can become much larger and are only bounded by the total number of players in this spatially unconstrained model. 
We find that high cooperation regimes exist 
for a wide range of parameters (see Figure~\ref{fig:PhaseDiagram}) as they did for the constrained system.

Intuitively, it makes sense that cooperative regions would be more stable and persist in the cellular automaton model once they develop, since such clusters are spatially segregated from defectors, but it is surprising that cooperation survives even in the spatially unconstrained model given that cooperators are always exposed to exploitation by invading defectors.

Our simulations show that cooperative behavior is the dominant strategy for certain parameter regions, regardless of initial conditions. In particular we observe the emergence of cooperation even when we start with a highly disadvantageous initial strategy distribution, where all players start out as defectors.

While the behavior of a network PGG shares qualitative similarities with the cellular automaton PGG, we also observe marked differences: parameter regions for cooperativity differ, relaxation times tend to be much longer for the network model, and state variable fluctuations (e.g., for instability, agglomeration, and cooperation) are much more pronounced in the network PGG where single game sizes are larger on average.


Observing the temporal development of system state variables like payoff, satisfaction, instability, agglomeration, and cooperation, we can distinguish several characteristic periods; see Figure~\ref{fig:3fields}. 
In each of those periods, state variables can fluctuate greatly over short time spans. 
In order to get a clearer picture of overall system trends, we primarily study moving average time series as in Figure~\ref{fig:stoch_coop}.

Starting from a network without player connections, consisting only of defectors, we find an initial period of low cooperation and agglomeration undergoes a phase transition into a late period of higher cooperation and agglomeration for a wide regime of parameters. 
Both periods begin rapidly and are marked by a high degree of instability.
In the later period, while the moving average number of cooperators is quite large, short-term fluctuations occur where the number drops to nearly zero.
This indicates strong global synchronization of player strategies made possible through the long-range interactions on the network.
Temporal variations of state variables of this magnitude are nonexistent for the constrained 2D lattice system.


The decision rule responsible for evolution in the model is based on player satisfaction.
Satisfied players do not change strategies or connections while unsatisfied players may change with nonzero probability.
However, an unsatisfied player may break a connection to a satisfied player so being satisfied does not guarantee stasis, even short-term.
The exponential decay of satisfaction due to habituation triggers the extreme fluctuations in payoff and cooperator fraction as well as network rearrangements as recorded by variables instability and agglomeration.
When the same habituation model is used on a 2D lattice, the effect of decaying satisfaction is significantly more localized due to players being limited to eight neighbors.


Despite dramatic temporal dynamics, characterization of typical long-term states is possible.
While a steady state on an individual player level has not been reached by the end of simulations after 20,000 time steps, heatmaps of the cooperator fraction and average player connections vary only slightly over the last several thousand steps; see Figures~\ref{fig:PhaseDiagram} and~\ref{fig:timed_phase_diagrams}.

We describe the PGG by a network model of interacting players. 
This description makes it easy to describe the characteristics of our particular network in a generic way.  
In particular, we are able to study the evolution of cooperation along with the growth of connections or edges across the network. 

We investigate the influence of different initial network topologies on final state topology and cooperation. 
Specifically, we compare starting from networks with no connections to Bar\'abasi-Albert networks generated by preferential attachment.\cite{Barabasi_1999,Albert_2002}
Initial player connections do not persist for very long and PGG cooperation dynamics are fairly universal after a transient period of network reconfiguration.
In all cases we find long time development towards random regular networks. 

The total number of edges grows during the simulations even though adding and removing edges at a certain time occurs with equal probability.  
Evidently, adding edges increases satisfaction more than removing edges. 
Indeed, a connection to a defector may actually increase a player's net payoff if that defector is connected to sufficiently many cooperators.

Edges are chosen for addition or removal at random and we observe the accumulation of edges among players of both types tends toward the expectation for the given number of cooperators and defectors; See Figures~\ref{fig:intraedges} and~\ref{fig:edgesCD}.
This is further illustrated by the full network degree distribution which becomes more narrow and peaked as time increases; see Figure~\ref{fig:degree-dist}.
Cooperation is robust to the random rewiring in the model and does not require the presence of high degree hubs to be maintained.
Notably, the long-term network topology is quite homogeneous with no scale-free statistics, even when initialized as a Bar\'abasi-Albert network.
This result is especially surprising in view of the fact that a fixed fully-connected network, homogeneous in every way, does not support cooperation, yet our cooperative PGG networks evolve toward such a network.

\EPtodo{I will concentrate on this section first over the weekend}

We suggest that our network PGG model is reminiscent of and can be used to describe several real life systems in which public goods are managed and exchanged like in open peer-to-peer networks, source code development groups, wikipedia/collaborative online work groups, crowd-source problem solving teams, insurance and loan markets.
It is important to notice that those systems might still be very fluid and undergo frequent and big changes on an individual level even if they operate in an overall cooperative regime.
\EPtodo{cancer treatment (shall we tie that in)}
\EPtodo{need references for these systems}

Future research will explore the influence of more realistic satisfaction measures, like the ones discussed in Kahneman\cite{Kahneman_2011} (prospect theory). \AStodo{Finish}
It might be possible to decouple single player behavior by allowing for limited options of changes at a certain time step (new connection formation, or severance of existing connections, or strategy changes  at a single time step instead of allowing all at the same time).\AStodo{Or, limit new edges to neighbors of neighbors only.}{} Similarly, asynchronous (vs. synchronous) updating might alleviate some synchronization effects and should be investigated.

\appendix

\section{Time series oscillations}\label{app:oscillations}
\begin{figure*}[!bh]
  \subfloat[]{\includegraphics[width=.4\textwidth]{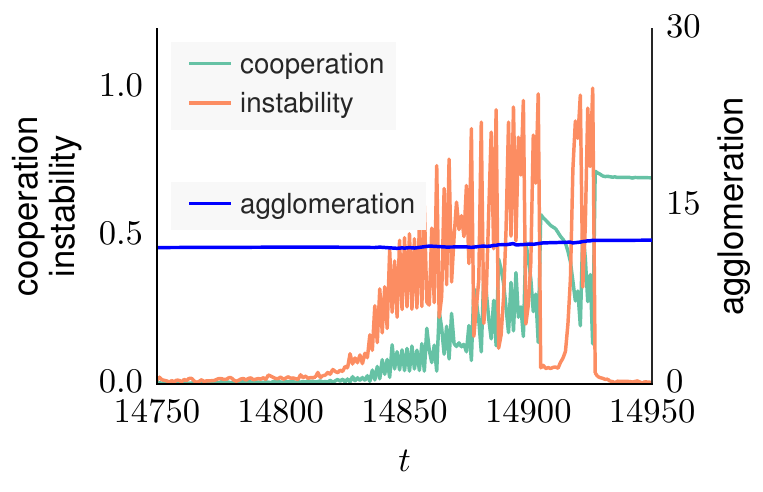}\label{fig:no_ave_zoom1}\quad}
  \subfloat[]{\includegraphics[width=.4\textwidth]{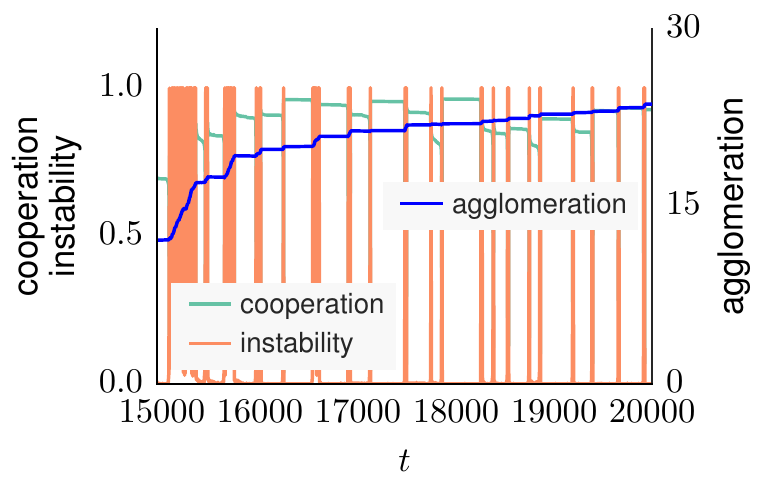}\label{fig:no_ave_zoom2}}
  \caption{Time series data without windowed averaging. Note the differing time axis scales in~\ref{fig:no_ave_zoom1} and~\ref{fig:no_ave_zoom2}. Parameters are as in Figure~\ref{fig:3fields}.}
  \label{fig:oscillations}
\end{figure*}

Cooperation, instability, and agglomeration time series are shown in Figure~\ref{fig:3fields} with a moving average over 100 time steps to illuminate long time trends. 
In Figure~\ref{fig:oscillations}, we plot the original data revealing the high frequency oscillations discussed in Section~\ref{sec:results} and the relationship between cooperation and instability.
In Figure~\ref{fig:no_ave_zoom1}, notice that instability involves mainly strategy changes toward cooperation yet in Figure~\ref{fig:no_ave_zoom2}, over a longer time frame, instability coincides with strategy changes as well as increasing agglomeration.

\section{Habituation rate derivation}\label{app:habituation}
In Section~\ref{sec:results}, we state that for short periods of time, a player's satisfaction decays exponentially at rate $\mu$. 
Here we show that explicitly by deriving Equation~\eqref{eq:satisfaction_decay}.
As before, assume $\pi_{i}(t)\approx\pi_{i}(t_{0})$ and $\pi_{i,\text{min}}(t) < \pi_{i}(t) < \pi_{i,\text{max}}(t)$ for \mbox{$t_{0}\le t\le t_{1}$}.
Then, from Equation~\eqref{eq:max_update} 
\begin{gather*}
\begin{aligned}
\pi_{i,\text{max}}(t\!+\!1) &= \pi_{i}(t) + (1\!-\!\mu)\left( \pi_{i,\text{max}}(t)-\pi_{i}(t) \right)\\
\pi_{i,\text{max}}(t\!+\!1) - \pi_{i}&(t_{0}) \\&\approx (1\!-\!\mu)\left[ \pi_{i,\text{max}}(t)-\pi_{i}(t_{0}) \right]\\
&\approx (1\!-\!\mu)^{2}\left[ \pi_{i,\text{max}}(t\!-\!1)-\pi_{i}(t_{0}) \right]\\
&\approx (1\!-\!\mu)^{3}\left[ \pi_{i,\text{max}}(t\!-\!2)-\pi_{i}(t_{0}) \right]\\
 \shortvdotswithin{\approx}
&\approx (1\!-\!\mu)^{t+1-t_{0}}\left[ \pi_{i,\text{max}}(t_{0})-\pi_{i}(t_{0}) \right],
\end{aligned}
\end{gather*}
where we've used $\pi_{i}(t)\approx\pi_{i}(t_{0})$. So, replacing $t+1$ by $t$, we find
\begin{align*}
  \pi_{i,\text{max}}(t) &\approx  (1\!-\!\mu)^{t-t_{0}}\left[ \pi_{i,\text{max}}(t_{0})-\pi_{i}(t_{0}) \right] + \pi_{i}(t_{0}).
\intertext{Similarly, by Equation~\eqref{eq:min_update}}
  \pi_{i,\text{min}}(t) &\approx  (1\!-\!\mu)^{t-t_{0}}\left[ \pi_{i,\text{min}}(t_{0})-\pi_{i}(t_{0}) \right] + \pi_{i}(t_{0}).
\end{align*}
We easily compute the aspiration and satisfaction from Equations~\eqref{eq:aspir} and~\eqref{eq:satis}, respectively:
\begin{align*} 
  \phantom{a_{i}(t) \approx}
  &\begin{aligned}
    \mathllap{a_{i}(t) \approx}\; 
    \pi_{i}(t_{0}) &- (1\!-\!\mu)^{t-t_{0}}\left[\pi_{i}(t_{0})- \pi_{i,\text{min}}(t_{0}) \right] \\
    & + \alpha (1\!-\!\mu)^{t-t_{0}}\left[ \pi_{i,\text{max}}(t_{0}) - \pi_{i,\text{min}}(t_{0}) \right]
  \end{aligned}\nonumber\\
  &\begin{aligned}
    \mathllap{=}\;
    \pi_{i}(t_{0}) - (1\!-\!\mu)^{t-t_{0}}\bigl(\!&\left[\pi_{i}(t_{0})- \pi_{i,\text{min}}(t_{0}) \right] \\
    &+ \alpha\! \left[ \pi_{i,\text{max}}(t_{0}) - \pi_{i,\text{min}}(t_{0}) \right] \bigr),
  \end{aligned}
\end{align*}
\vspace{-10pt}
\begin{equation*}
\begin{aligned}
  s_{i}(t) \approx  (1\!-\!\mu)^{t-t_{0}}\bigl(&\left[\pi_{i}(t_{0})- \pi_{i,\text{min}}(t_{0}) \right] \\
  &- \alpha \left[ \pi_{i,\text{max}}(t_{0}) - \pi_{i,\text{min}}(t_{0}) \right] \bigr) + \eta_{i}(t).
\end{aligned}
\end{equation*}
Since $(1-\mu)^{t-t_{0}} = \exp\left((t-t_{0})\log(1-\mu)\right)$, satisfaction decays exponentially at a rate $\left|{\log(1-\mu)}\right| \approx \mu$ for $0<\mu\ll1$ as given by Equation~\eqref{eq:satisfaction_decay} in Section~\ref{sec:results}.

\section{Cooperation phase diagrams over time}
\begin{figure*}[!hbp]
  \subfloat[$t=100$]{\includegraphics[width=.3\textwidth]{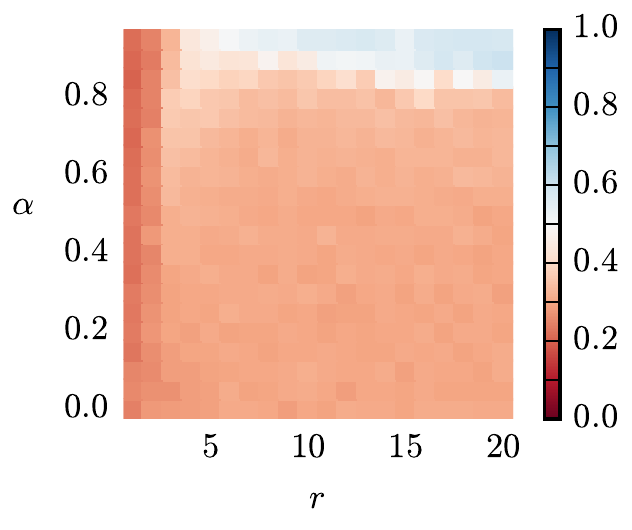}}\quad
  \subfloat[$t=250$]{\includegraphics[width=.3\textwidth]{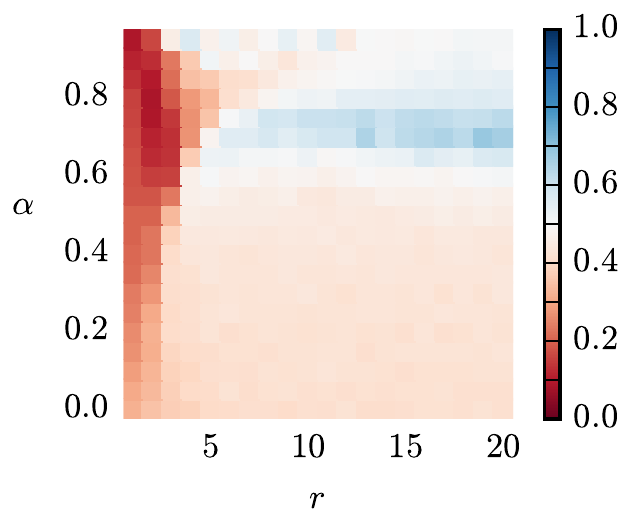}}\quad
  \subfloat[$t=500$]{\includegraphics[width=.3\textwidth]{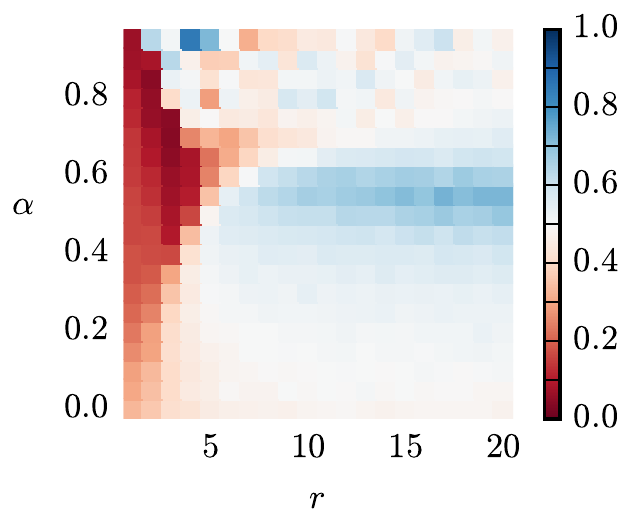}}\\
  \subfloat[$t=1000$]{\includegraphics[width=.3\textwidth]{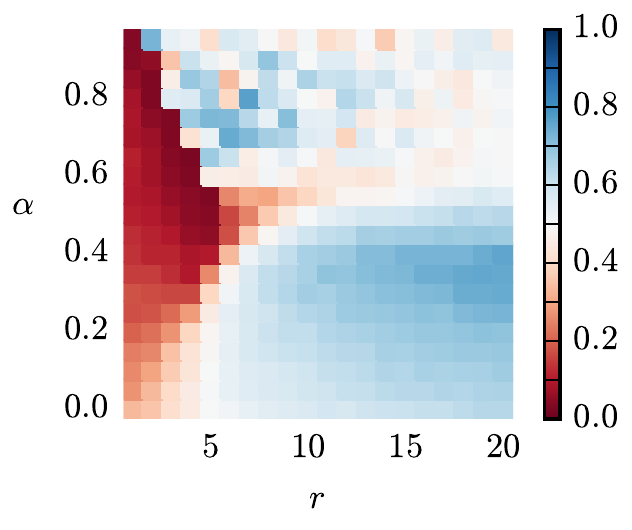}}\quad
  \subfloat[$t=10000$]{\includegraphics[width=.3\textwidth]{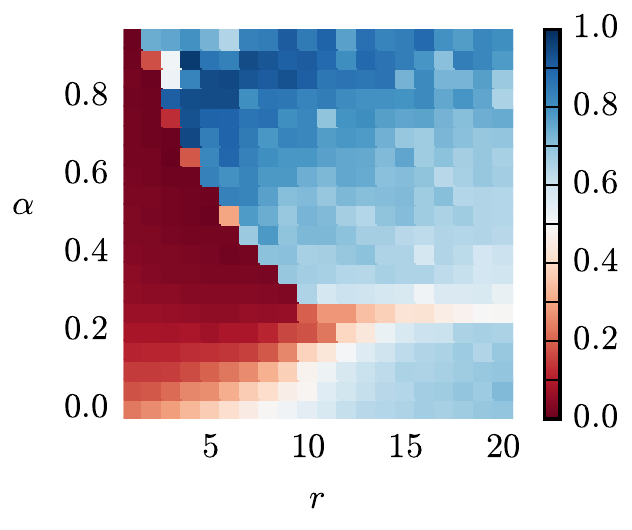}}\quad
  \subfloat[$t=20000$]{\includegraphics[width=.3\textwidth]{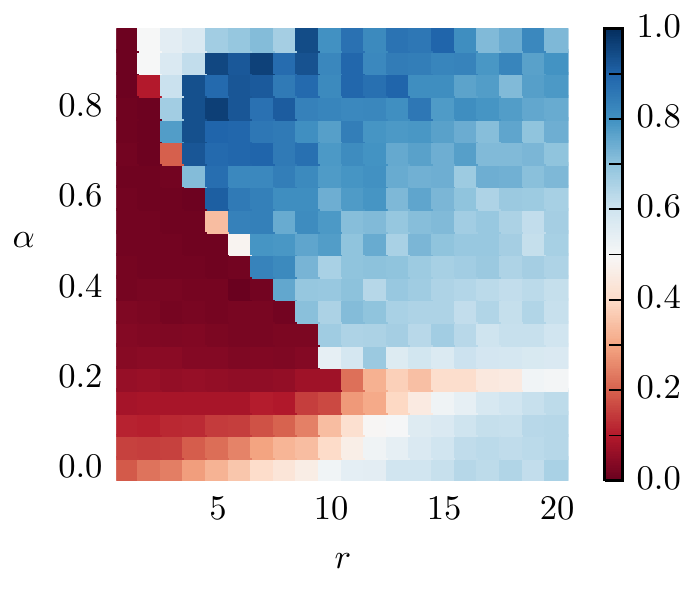}}\\
  \caption{Phase diagram of cooperation at different simulation time steps. Parameters are as in Figure~\ref{fig:PhaseDiagram}.
  \label{fig:timed_phase_diagrams}}
\end{figure*}

The phase diagram in Figure~\ref{fig:PhaseDiagram} reveals the states of systems at $t=20000$. 
Since simulations for various parameters reach their characteristic long term states at different times, it is interesting to observe the phase diagram develop over time, as in Figure~\ref{fig:timed_phase_diagrams}. Note in particular the similarity between $t=10000$ and $t=20000$. 

\clearpage 

\bibliographystyle{abbrv}
\bibliography{../../Literature/bibliography}

\end{document}